\documentclass[aps,prl,twocolumn,superscriptaddress,floatfix,showpacs]{revtex4-1}

\usepackage{graphicx}
\usepackage{amsmath}

\begin{document}

\title{Disentangling factors governing Dzyaloshinskii domain wall creep in Co/Ni thin films using Pt$_x$Ir$_{1-x}$ seedlayers}

\author{D. Lau}

\author{J.P. Pellegren}

\affiliation{Department of Materials Science \& Engineering, Carnegie Mellon University, Pittsburgh, Pennsylvania 15213, USA}

\author{H.T. Nembach}
\author{J.M. Shaw}

\affiliation{Quantum Electromagnetics Division, National Institute of Standards and Technology, Boulder, CO 80305, USA}

\author{V. Sokalski}
\email[]{vsokalsk@andrew.cmu.edu}
\affiliation{Department of Materials Science \& Engineering, Carnegie Mellon University, Pittsburgh, Pennsylvania 15213, USA}

\date{\today}

\begin{abstract}

	We characterize asymmetric growth of magnetic bubble domains in perpendicularly magnetized Co/Ni multi-layers grown on Pt$_x$Ir$_{1-x}$ seedlayers by application of perpendicular and in-plane magnetic fields.  Using a refined model of domain wall creep that incorporates contributions from the anisotropic elastic energy, $\varepsilon$, and a chirality-dependent prefactor, $v_0$, we elucidate factors that govern the mobility of Dzyaloshinskii domain walls as a function of seedlayer composition.  The interfacial Dzyaloshinskii-Moriya Interaction magnitude is found to decrease monotonically with $x_{Ir}$, which is independently confirmed by Brillouin light scattering (BLS).  Moreover, the persistence of significant asymmetry in velocity curves across the full composition range supports previous assertions that a chirality-dependent attempt frequency akin to chiral damping could play a critical role in the observed trends.  This work helps resolve fundamental questions about the factors governing Dzyaloshinskii DW creep and demonstrates varying Pt-Ir seedlayer composition as a method to tune DMI.  

\end{abstract}

\maketitle

Recent observations that topologically protected magnetic features like skyrmions and chiral domain walls (DWs) can be manipulated with spin current has renewed interest in developing spintronic devices for energy efficient nonvolatile memory and logic applications \cite{7047159,Emori_2013,Ryu_2014,Woo_2016,PhysRevB.78.140403,Heinze:2011aa}. These topological structures are stabilized by the Dzyaloshinskii-Moriya Interaction, DMI, which is an anti-symmetric exchange energy that scales as $E=-\textbf{D}\cdot (\textbf{S}_1\times \textbf{S}_2)$ leading to chiral winding configurations as the ground state.\cite{DZYALOSHINSKY1958241,PhysRev.120.91} Here, $\textbf{S}$ represents the spin angular momentum of neighboring electrons and $\textbf{D}$ is the DMI vector.  Prospects for future thin film engineering in this area were bolstered by the discovery of an interfacial DMI, iDMI, that exists in ultrathin heavy metal / ferromagnet heterostructures because of their structural inversion asymmetry (SIA).\cite{0295-5075-100-5-57002} In this case, $\textbf{D}$ is restricted to lie in the plane of the film with direction given by $\textbf{D} = D(\hat{r}\times\hat{z})$ where $\hat{r}$ and $\hat{z}$ are the unit vectors from $\textbf{S}_1$ to $\textbf{S}_2$ and the film normal, respectively. The impact of several seedlayers and their thickness have been explored experimentally in an effort to control the strength of this effect.\cite{Hrabec_2014,PhysRevB.88.214401,Nembach:2015aa,Soucaille_2016,0953-8984-27-32-326002}  However, to date there have only been theoretical investigations on the composition dependence of the iDMI, which we present here for Pt-Ir alloys.\cite{2017arXiv171102657H}  In thin films with a perpendicular magnetization, $\textbf{D}$ can be described by an effective field, $\mu_oH_{DMI} = D/(M_s\lambda)$, that acts on the internal magnetization of a DW favoring the N\'{e}el configuration over the in the out-of-plane geometry magnetostatically favored Bloch type, where $M_s$ and $\lambda$ are the saturation magnetization and Bloch wall width, respectively.\cite{Chen2013,Hrabec_2014,PhysRevB.88.214401}   It is now well-established that the combination of $H_{DMI}$ and an in-plane field $H_x$ leads to a wall energy that is highly anisotropic with respect to the DW normal's spatial orientation about $H_x$. \cite{Lau_2016,doi:10.1063/1.4881778,Pellegren:2016aa}  This break in symmetry results in asymmetric expansion of magnetic bubble domains when subjected to a perpendicular driving field.\cite{Kabanov2010, Hrabec_2014,PhysRevB.88.214401,Lavrijsen_2015}  For small driving fields, the motion is thermally activated with velocity described by the Arrhenius creep scaling law, $v = v_o e^{\zeta H_z^{-1/4}}$, where $\zeta$ has built in the activation energy for DW propagation and is proportional to the fourth root of the DW elastic energy, $\varepsilon^{1/4}$. The prefactor, $v_0$, is the corresponding attempt frequency for DW propagation.\cite{Blatter1994,Lemerle1998}  Although asymmetric domain growth has become the predominant technique for extracting $D$, fundamental questions remain about how to interpret creep velocity changes with $H_x$ in ultrathin ferromagnetic films with appreciable iDMI.

\begin{figure}
	\includegraphics[width = 3.4in]{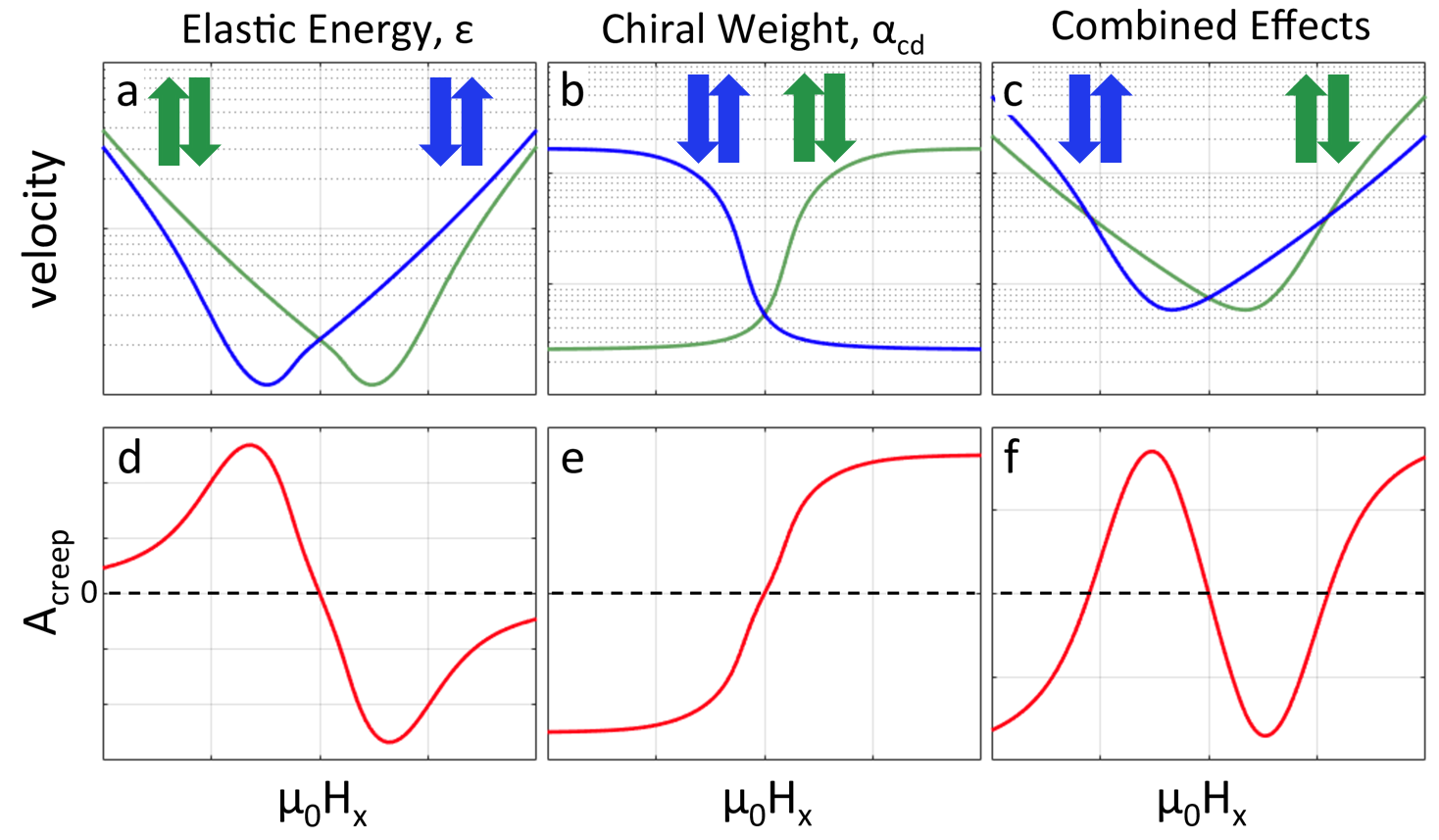}
	\caption{\label{stiff_cd_effects}Illustration of the impact of the anisotropic elastic energy (a) and chiral weight (b) on velocity vs $\mu_0H_x$ with their combined effect shown in (c).  d-f) The corresponding effects on A$_{creep}$. Most notable is the convergence of $\uparrow\downarrow$ (blue) and $\downarrow\uparrow$ (red) domain walls at large $\mu_0H_x$ for the case of elastic energy alone.  This convergence is absent for a non-zero chiral weight.} 
\end{figure}

\begin{figure*}
	\includegraphics[width = 7in]{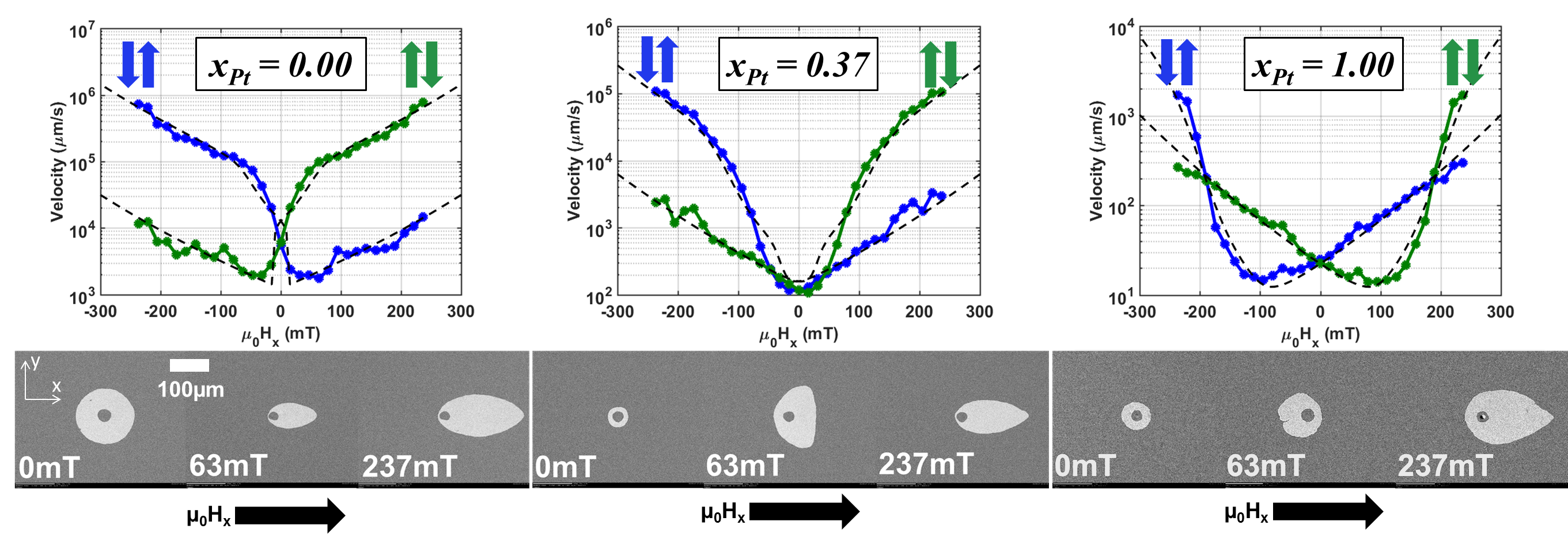}
	\caption{\label{repLNv}Experimental v vs. $\mu_0H_x$ for seedlayers with varying $x_{Pt}$ with representative MOKE images. Dashed lines are fits from equation \ref{eq:FullCreep}.  The center grey of the Kerr images represent the initial bubble shape while the white region is the domain after growth under both $\mu_0H_x$ and $\mu_0H_z$, which was fixed at 7 mT.}
\end{figure*}

Initial work on this topic suggested that $\varepsilon$ is equivalent to $\sigma$, the wall energy, and was the factor governing DW velocity.  Assuming constant $\lambda$, $\sigma$ vs. $H_x$ is symmetric about a maximum that occurs when $H_x = H_{DMI}$ and was proposed to correspond to a minimum in velocity.\cite{Hrabec_2014,PhysRevB.88.214401}  Significant asymmetric deviations from this idealized shape observed experimentally led to speculation about other possible factors that could be contributing.\cite{Akosa_2016,Ju__2015,1882-0786-9-5-053001,Lavrijsen_2015}  This included chiral damping, which would impact $v_0$ instead of $\varepsilon$ and depend only on the orientation of the DW internal magnetization.\cite{Ju__2015}   It was also later identified that $\varepsilon$ is actually given by the stiffness, $\tilde{\sigma}(\Theta) = \sigma(\Theta)+\sigma''(\Theta)$, which should reside in the exponent of the creep law instead of $\sigma(\Theta)$.\cite{Pellegren:2016aa}   Here $\Theta$ denotes the angle between the DW normal and $H_x$.  In the isotropic case, $\sigma''(\Theta)= 0$. However, in cases of anisotropic DW energy as found in Dzyaloshinskii DWs subject to $H_x$ here, $\sigma''(\Theta)$ becomes comparable in magnitude $\sigma(\Theta)$.  This description based only on elastic energy of the domain wall demonstrated that significant curve asymmetry should exist due exclusively to iDMI and was able to explain some of the perplexing experimental data. 

In this paper, we use an augmented model for Dzyaloshinskii DW creep to fit experimental measurements of asymmetric domain growth in Co/Ni multi-layers grown on Pt-Ir alloy seedlayers.  The model incorporates elastic energy of the domain wall based on its dispersive stiffness \cite{Pellegren:2016aa} and also allows for a chirality dependent prefactor that would occur in the case of chiral damping given by $v_0(H_x) = v_0^*(1+\alpha_{cd}cos(\phi_{eq}(H_x)-\Theta))$ where $\alpha_{cd}$ is a parameter from -1 to 1 that characterizes the weight of this effect - hereafter referred to as the chiral weight.  $\phi_{eq}$ is the equilibrium internal magnetization orientation with respect to $H_x$ as calculated in Pellegren et al.\cite{Pellegren:2016aa} $v_0^*$ is the attempt frequency absent any chiral effects.  For calculations of $v_0$, we only consider the case of $\Theta = 0$ or $\pi$ to account for fits to the left and right velocities of the bubble domains.  The resulting creep equation describing DW velocity as a function of $H_x$ is given as follows:

\begin{equation}\label{eq:FullCreep}{v=v_0(H_x)exp   \left [\kappa   \frac{\tilde{\sigma}(H_x)}{\tilde{\sigma}(H_x=0)}H_z^{-1/4}  \right ]}\end{equation}

	where $\kappa$ is a creep scaling constant that does not depend on $H_x$.  $\tilde{\sigma}$ is calculated using the dispersive stiffness model of Pellegren et al. in the limit of a vanishingly small deformation lengthscale, $L$ as justified later.\cite{Pellegren:2016aa}  The effects of elastic energy and chiral weight on the shape of velocity curves is shown qualitatively in Figure \ref{stiff_cd_effects}.  The asymmetric component, $A_{creep}=\ln(v_{\uparrow\downarrow}/v_{\downarrow\uparrow})$, is included to further highlight experimental signatures associated with the different mechanisms. $(v_{\uparrow\downarrow}$ and $v_{\downarrow\uparrow}$  are the domain wall velocities, where the magnetization transitions from up to down and down to up, respectively. In the case where only elastic energy is considered, $v_{\uparrow\downarrow}$ and $v_{\downarrow\uparrow}$ converge (i.e. $A_{creep} = 0$) as $H_x\to\infty$.  For a non-zero $\alpha_{cd}$, $A_{creep}$ saturates when $H_x > H_{DMI} + H_{DW}$ where $H_{DW}$ is the DW anisotropy field.

Co/Ni films were prepared using DC magnetron sputtering from 5 in. targets onto 3 in. Si (001) substrates with native oxide.  The working pressure was fixed at 2.5 mTorr Ar. The film stack is Substrate/TaN(3)/Pt(3.5)/Pt$_{x}$Ir$_{1-x}$(1.2)/ [Co(0.2)/Ni(0.6)]$_2$/Co(0.2)/Ta(0.8)/TaN(6), with units in nanometers. The Pt$_{x}$-Ir$_{1-x}$ seedlayer is prepared using a combinatorial sputtering technique where the substrate is moved between two targets rapidly, depositing $<$ 0.05 nm of material in each cycle to mimic the cosputtering process. This results in a linear composition gradient across the substrate surface. Details on the structural characterization of similar Co/Ni multi-layer films can be found in \cite{doi:10.1063/1.4982163}. $M-H$ loops measured using alternating gradient field magnetometry (AGFM) and vibrating sample magnetometry (VSM) across the composition gradient indicate a saturation magnetization, $M_{s} \sim 645 kA/m$, and in-plane saturation field, $\mu_0H_k \sim1.3T$, which has little dependence on Pt$_x$Ir$_{1-x}$ seedlayer composition (see supplemental information (S1)).  Measurement of domain growth was performed using a wide-field white light Kerr microscope. The microscope is fit with an in-plane electromagnet capable of producing static in-plane fields up to $\mu_0H_x$ $\sim$ 250 mT as well as a perpendicular coil that can generate up to $\mu_0H_p$ $\sim$ 20 mT magnetic pulses down to 1 ms. As described in \cite{Lau_2016,Pellegren:2016aa}, a $Ga^{+}$ ion beam is used to selectively damage portions of a sample film, where initial bubble domains of approximately 20$\mu$m can be nucleated. Velocity was determined by two images showing the difference in domain wall positions before and after a single pulse.  The pulse length ranged from 1-20ms and was chosen so that an appreciable displacement would occur. 

We used Brillouin Light Scattering spectroscopy (BLS) to establish an independent measure of the magnitude of the DMI. The laser had a wavelength of 532 nm. Damon-Eshbach spin-waves experience a non-reciprocal frequency-shift $\Delta f_{DMI}=|\frac{g^{||}\mu_B}{h}|sgn(M)\frac{2D}{M_s}k $ in the presence of DMI. The spectroscopic splitting factor is estimated as $g^{||} = 2.19$ \cite{Arora2017}, $ \mu _B$ is the Bohr Magneton, h is Planck's constant and $\textbf{k}$ is the spin-wave wavevector with $|k|=16.7 \mu m^{-1}$. We measured the spin-wave frequency for the two opposite directions of the magnetization to determine $\Delta f_{DMI}$. The measured $|\Delta f_{DMI}|$ was between 0.1 GHz and 0.8. GHz

\begin{figure}
	\includegraphics[width = 3.1in]{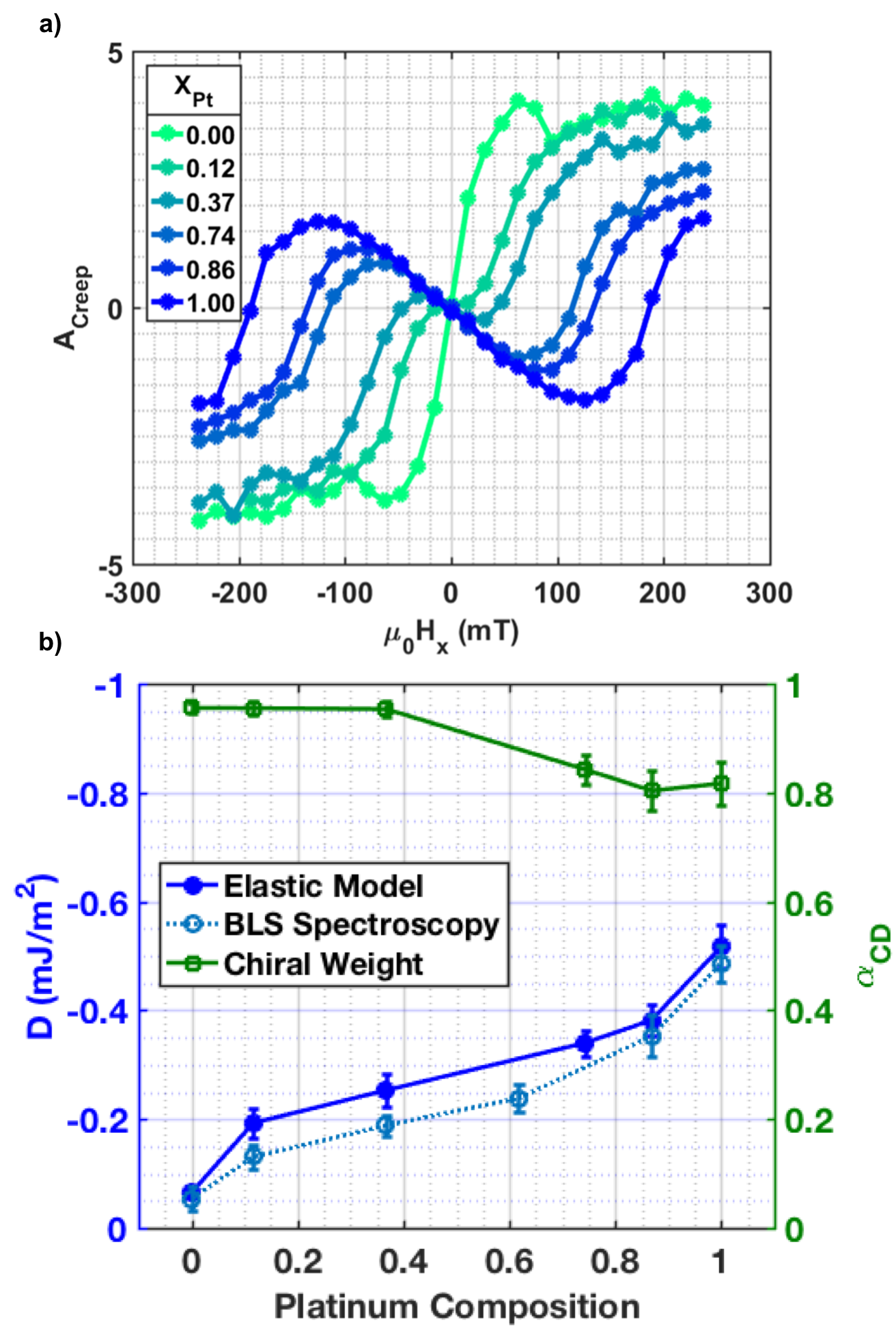}
	\caption{\label{dmiasymmetry}a) A$_{creep}$ vs. $\mu_0H_x$ as a function of X$_{Pt}$. b) Extracted values of $D$ and $\alpha_{cd}$ vs X$_{Pt}$ based on fits to $A_{creep}$. Closed blue (dark) circles represent fits extracted from the elastic domain wall model. Open blue (light) circles represent $D$ values characterized using BLS demonstrated by Nembach et al.\cite{Nembach:2015aa}}
\end{figure}

Figure \ref{repLNv} shows representative Kerr images as a function of in-plane field ($H_x$) and seedlayer composition with corresponding $v$ vs $H_x$ curves.  As seen in previous studies, the domain shape is highly non-elliptical evolving from a flattened shape at low field to a teardrop shape at higher field.\cite{Lau_2016,Lavrijsen_2015,Woo2016}  The field at which this occurs is found to be directly related to the amount of Pt in the seedlayer (see supplementary info for additional Kerr images).  To separate the effects of elastic energy and chiral weight, we examine the shape of $v$ vs $H_x$ and the calculated $A_{creep}$ (Figure \ref{dmiasymmetry}).  In all cases, the velocity curve is asymmetric about a minimum in velocity.  This leads to a reversal in the preferred expansions direction in the Pt-rich compositions, which is indicated by the intersection of the velocity curves in figure 2 and by the zero crossing of $A_{creep}$ in figure 3 a. As identified previously, a change in sign of A$_{creep}$ at non-zero $H_x$ could be explained using a larger deformation lengthscale, L, in the dispersive stiffness model.\cite{Pellegren:2016aa}  However, the observation that A$_{creep}$ tends to saturate rather than return to zero suggests this is not the case.  Therefore, we limit our fitting to the case of L $\to0$, which is consistent with the expectation that pinning sites in sputtered thin films are densely distributed.  We note that as the composition shifts from $x_{Pt} = 1$ to $0$, the minimum in velocity shifts towards $H_x=0$ and changes sign near $x_{Pt} = 0.25$.   However, as the creep fits and BLS measurements show, $D$ does not actually change sign and only approaches 0 for the case of pure Ir.  This result is in stark contrast to the aforementioned creep models based only on the wall energy, which would have given the incorrect sign of $D$ in this range.\cite{Hrabec_2014,PhysRevB.88.214401}  

Even as $D$ decreases with decreasing $x_{Pt}$, the asymmetry of the curve is preserved suggesting that its origin is not exclusively due to iDMI.  Indeed, $A_{creep}$ appears to saturate in all cases even though its magnitude is reduced for increasing Pt content.  The results of the fit to the velocity curves are shown in Figure \ref{dmiasymmetry}b highlighting that significant $\alpha_{cd}$ is needed to explain the data of Figures \ref{repLNv}/\ref{dmiasymmetry}a and dominates the trend for large $X_{Ir}$.

\begin{figure}
	\includegraphics[width = 3.4in]{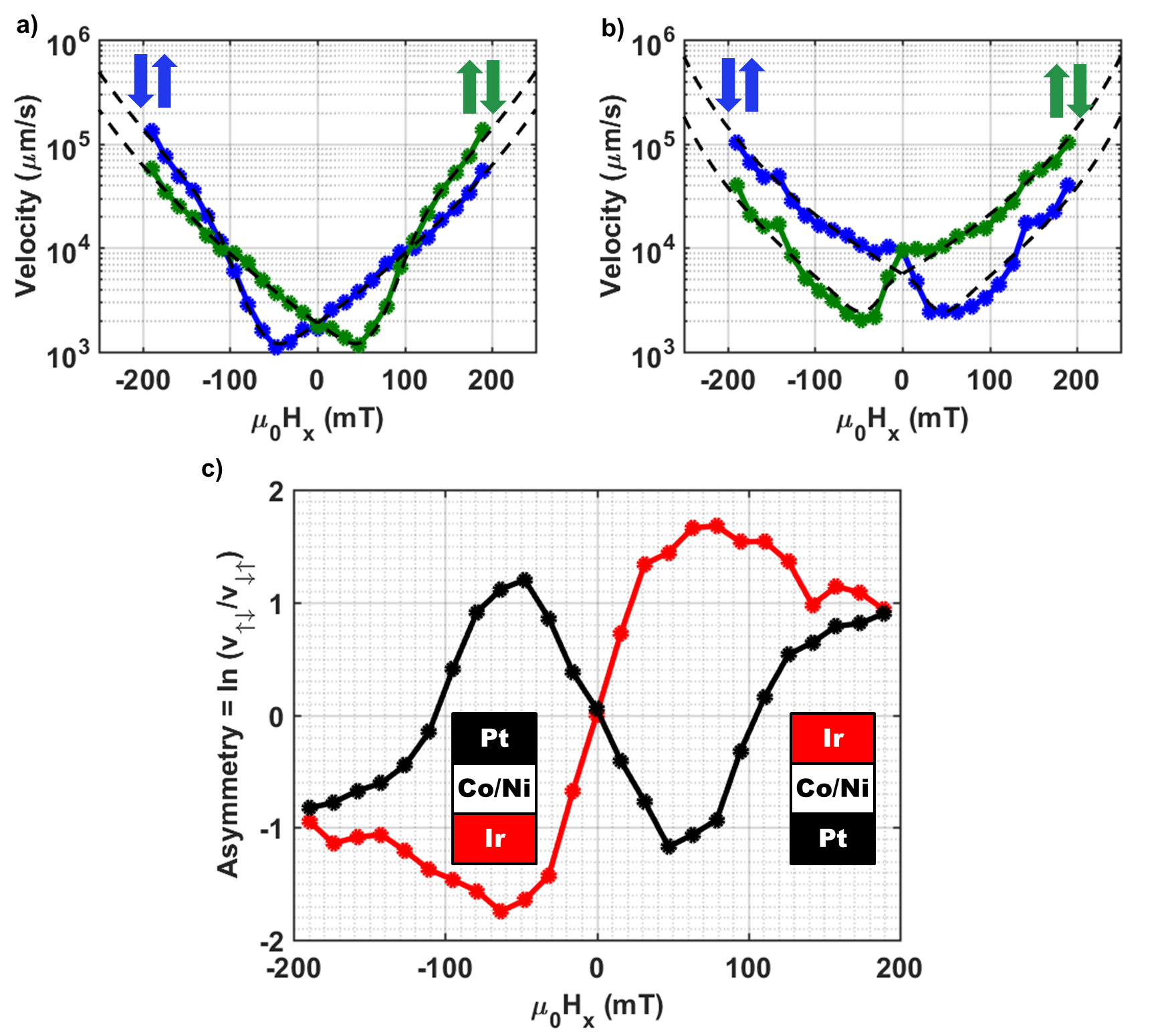}
	\caption{\label{alt_films}Experimental velocity vs. $\mu_0H_x$ for samples grown with a) Pt seedlayer/Ir capping layer and b) Ir seedlayer/Pt capping layer. c) $A_{creep}$ vs $\mu_0H_x$ calculated from experimental velocity data in (a) and (b).} 
\end{figure}

	To further examine the impact of chiral weight and iDMI via the elastic energy, we have prepared the following films: TaN(3)/Pt(2.5)/[Co(0.2)/Ni(0.6)]$_2$/Co(0.2)/Ir(2.5)/TaN(6) and the same stack with Pt and Ir positions swapped.  These are referred to as Pt-seed/Ir-cap and Ir-seed/Pt-cap, respectively.  Velocity curves and asymmetry for these samples are shown in Figure \ref{alt_films}.  We note that the magnitude of $D$ measured here should not be compared with the results tabulated in Figure \ref{dmiasymmetry} because we have significantly increased the effective magnetic layer thickness by replacing the Ta cap (known to create a magnetic dead layer) with either Pt or Ir (both known to have a proximity induced magnetization).  Indeed, we see that the sign of $D$ is reversed between these two cases with comparable magnitudes as expected.  The Pt seed/Ir cap favors left-handed N\'{e}el walls ($D = -0.313\pm0.009 mJ/m^2$) and the Ir seed/Pt cap favors right-handed N\'{e}el walls ($D = 0.214\pm0.020mJ/m^2$).  It is interesting that despite the expected change in sign of $D$, $\alpha_{cd}$ remains nearly the same ($\alpha_{cd,Pt-seed}=0.41\pm0.04$, $\alpha_{cd,Ir-seed}=0.58\pm0.07$).  If $\alpha_{cd}$ depended exclusively on the elements present and interface orientation, we should see a change in sign upon reversal of the film stack.  The absence of this reversal suggests that there could be a contribution to the chiral weight that is intrinsic to the Co/Ni stack even though it is nominally symmetric.  Just as Pt/Co/Pt films are known to have SIA, it is conceivable that the Co/Ni/Co/Ni/Co film stack itself could be structurally asymmetric if the lattice evolves through the thickness and/or the top and bottom Co/Ni interfaces are not identical.  This assertion requires further investigation as it is also possible that the chiral weight contributions from Pt and Ir change when used as seed vs cap layers.

In summary, we have shown a monotonic increase of $D$ with X$_{Pt}$ in Pt$_x$Ir$_{1-x}$ seedlayer alloys.  Moreover, we show that the impact of DMI on elastic energy is insufficient to explain the trends in velocity curves seen experimentally.  The results are fit well when a chirality-dependent attempt frequency is included in the model --- something speculated to originate from chiral damping or, more recently, a chiral gyromagnetic ratio.\cite{Ju__2015,Kim2018}  However, it remains unclear if the 10-100x increase in velocity is consistent with these mechanisms.  We also show definitively that reversal of Pt and Ir stack sequence indeed reverses the sign of $D$, but does not change the sign of $\alpha_{cd}$. This suggests that there could be a mechanism for chiral effects built into the Co/Ni multi-layers themselves.  The ability to tune iDMI via Pt-Ir composition and through reversal of Pt:Ir stacking sequence as demonstrated here provides new guidance for the design of film stacks in future spintronic applications.

This work has been funded by the DARPA Topological Excitations in Electronics (TEE) program and also funded (in part) by the Dowd Fellowship from the College of Engineering at Carnegie Mellon University. The authors would like to thank Philip and Marsha Dowd for their financial support and encouragement as well as Emilie Ju\'{e} for her help with the BLS data analysis. 

\bibliography{pt-ir}

\end{document}